# Magnetic, structural and magnetocaloric properties of $Y_{0.9}Gd_{0.1}Fe_2H_x$ hydrides

V. Paul-Boncour [1*], K. Provost[1], T. Mazet[2], A. N'Diaye[1], E. Alleno[1], F. Couturas[1]

[1]Univ. Paris-Est Créteil, CNRS, ICMPE, UMR7182, F-94320 Thiais, France
[2]Université de Lorraine, CNRS, IJL, F-54000 Nancy, France

**Abstract**

At 300 K, $Y_{0.9}Gd_{0.1}Fe_2H_x$ hydrides crystallize sequentially with increasing H concentration in various structures related to a lowering of the cubic $MgCu_2$ type structure of the parent alloy: cubic *C1*, monoclinic *M1*, cubic *C2*, monoclinic *M2*, cubic *C3*, orthorhombic *O*. Above 300 K, they undergo a first-order transition at a $T_{O-D}$ temperature driven by order-disorder of hydrogen atoms into interstitial sites. Their magnetic, structural and magnetocaloric properties have been investigated through magnetic measurements, and high-resolution synchrotron diffraction experiments. The magnetization at 5 K decreases slightly from 4 to 3.8 $\mu_B$ for $x = 3$ to 3.9 H/f.u., then with a larger slope for higher H content. A discontinuous decrease of the magnetic transition temperature is observed: *M1* and *C2* hydrides are ferrimagnetic with $T_C$ near 300 K, *M2* hydride displays a sharp ferromagnetic-antiferromagnetic transition at $T_{FM-AFM}$ =144 K, whereas *C3* and *O* hydrides present only a sharp increase of the magnetization below 15 K and a weak magnetization up to RT. Negative magnetic entropy variations ($\Delta S_M$) are measured near $T_C$ for the *M1* and *C2* phases, near $T_{FM-AFM}$ for the *M2* phase, whereas positive $\Delta S_M$ peaks due to inverse MCE effect are found near $T_{O-D}$. A structural and magnetic phase diagram is proposed.

## 1-Introduction

Development of efficient magnetocaloric materials (MCM) has become challenging since the discovery of a giant magnetocaloric effect near room temperature (RT) in $Gd(Ge,Si)_5$ compounds [1, 2]. Several other materials with giant magnetocaloric effect [3] such as $La(Fe,Si)_{13}$ [4, 5], MnFePSi [6-9], MnAs [10, 11] have been found and widely studied since this time. These giant MCE are associated to first order magnetic transitions and often accompanied by structural transitions.

$R$Ni$_2$ and $R$Co$_2$ Laves phase intermetallic compounds ($R$ = rare earth) present also giant MCE at low temperature near their magnetic ordering transition temperatures [12-17]. But the MCE effect decreases significantly upon heating. An alternative to find MCE near RT in Laves phase compounds is to decrease the magnetic ordering temperature in $R$Fe$_2$ compounds by hydrogen insertion. $R$Fe$_2$ intermetallic compounds are either ferromagnetic ($R$ = Y, Pr, Nd, Lu) or ferrimagnetic ($R$= Gd to Tm) with Curie temperature $T_C$ around 600-700 K [18-20]. Hydrogen insertion in YFe$_2$ induces a decrease of $T_C$ and a slight increase of the Fe moment up to 3.5 H/f.u. [21, 22]. For $x$ = 4.2 H/f.u. a ferromagnetic-antiferromagnetic (FM-AFM) transition has been observed at $T_{\text{FM-AFM}}$ and an itinerant electron metamagnetic (IEM) behavior found above this transition [23]. YFe$_2$H$_5$ behaves as a Pauli paramagnet with a small spontaneous magnetization of 0.4 µ$_B$/f.u. [24]. According to band structure calculations the ground state magnetization of the hydrides results from a competition between two antagonist effects: the lattice expansion which increases the Fe magnetic moments and the Fe-H interactions which reduce them [25-27]. YFe$_2$H$_5$, which contains the largest H content, is theoretically found more stable in a non-magnetic state in agreement with the experimental results [26].

A value of magnetic entropy variation ($\Delta S_M$) close to that of Gd has been observed in monoclinic YFe$_2$D$_{4.2}$ and YFe$_2$H$_{4.2}$ around $T_{\text{FM-AFM}}$ = 84 K and 131 K respectively due to the itinerant electron metamagnetic (IEM) behavior of the Fe sublattice [28]. $T_{\text{FM-AFM}}$ is highly sensitive to any volume change induced by applying an external pressure [29], H for D isotope substitution [30] or chemical substitution of Y by another rare earth element ($R$ = Gd, Tb, Er) [31-34]. To observe an MCE effect close to room temperature, we have undertaken a more systematic study of the influence of the chemical substitution for hydrides and deuterides with concentration near 4.2 H(D)/f.u.. The partial substitution of Y by an element with a larger metallic radius such as Gd allows to shift $T_{\text{FM-AFM}}$ to larger values and the transition temperature increases from 98 K for $Y_{0.9}Gd_{0.1}Fe_2D_{4.3}$ to 144 K for $Y_{0.9}Gd_{0.1}Fe_2H_{4.4}$. MCE was also observed in these compounds at $T_{FM-AFM}$ [35].

The transition temperatures remain however far from room temperature. To obtain hydrides with MCE near room temperature, we decided to investigate more systematically the influence of the H concentration on the magnetic and magnetocaloric properties of $Y_{0.9}Gd_{0.1}Fe_2H_x$ hydrides. For this purpose, several $Y_{0.9}Gd_{0.1}Fe_2H_x$ hydrides ($2.9 \leq x \leq 5$) have been synthetized and their structural properties studied as described in [36]. The hydrides crystallize at room temperature in cubic (*C1*, *C2* and *C3*), monoclinic (*M1* and *M2*) or orthorhombic (*O*) structures, with the following sequence *C1*, *M1*, *C2*, *M2*, *C3*, *O* for increasing H content and are separated by two phase ranges. The monoclinic and orthorhombic structures are obtained by a distortion of the $MgCu_2$ type cubic structure of the parent $Y_{0.9}Gd_{0.1}Fe_2$ compound, whereas the cubic *C2* phase displays a cubic $2a$ superstructure. The lowering of crystal symmetry can be attributed to a long-range order of hydrogen atoms into preferential interstitial sites. Upon heating the hydrogen long range order is lost and transitions towards a disordered cubic structure $C_{Dis}$ are observed leading to reversible order-disorder transitions at $T_{O-D}$ for *M1*, *C2* and *M2* [36].

In the present work the magnetic, structural and magnetocaloric properties of the $Y_{0.9}Gd_{0.1}Fe_2H_x$ hydrides are investigated. This is performed by combining magnetic measurements with high resolution X-ray powder diffraction measurements using synchrotron radiation (SR-XRD) versus temperature. The evolution of the magnetic properties and magnetic entropy variation versus H concentration is reported and compared with other materials.

## 2-Experimental methods

$Y_{0.9}Gd_{0.1}Fe_2$ intermetallic compound has been prepared by induction melting of the pure elements followed by three weeks annealing treatment under vacuum at 1100 K. The alloy composition checked by EPMA was $Y_{0.88(2)}Gd_{0.12(2)}Fe_{2.02(8)}$. It is single phase and crystallizes in a cubic $MgCu_2$ type structure with $a = 7.3638(5)$ Å and $V = 399.30(6)$ Å$^3$.

$Y_{0.9}Gd_{0.1}Fe_2H_x$ ($x < 4.5$) were prepared by solid-gas reaction as described in [37]. $Y_{0.9}Gd_{0.1}Fe_2H_{4.8}$ was prepared using a high-pressure device [24, 38]. The sample holders were quenched into liquid nitrogen while keeping the hydrogen pressure, then they were evacuated, filled with air, removed from liquid nitrogen, and kept heating slowly up to room temperature before removing the powdered hydride sample from the container. This procedure was found effective to poison the surface and prevent gas desorption.

The samples were characterized by XRD at room temperature with a Bruker D8 diffractometer ($\lambda_{K\alpha1}$ = 1.54059 and $\lambda_{K\alpha2}$ = 1.54438 Å). X-ray diffraction measurements using synchrotron radiation versus temperature were performed for selected samples on a 2-circle diffractometer of CRISTAL beam line at SOLEIL (Saint Aubin, France). A multi-crystals (21) analyser was used to obtain high angular resolution diagrams and the wavelength, refined with a LaB$_6$ (NIST 660a) reference sample, was $\lambda$ = 0.582644 Å. Additional measurements were made at higher energy ($\lambda$ = 0.51302 Å) using a 9 linear Mythen modules covering a 50° $2\theta$ circle arc at 720 mm from the sample detector. The powder samples were placed in sealed glass capillary tubes (0.3 mm diameter), which were rotated to ensure homogeneity. The samples were cooled or heated using a gas streamer cooler operating up to 380 K. All the SR-XRD patterns were refined with the Fullprof code [39].

Magnetic measurements were performed using a conventional Physical Properties Measurement System (PPMS) from Quantum Design with a maximum field of 9 T. The thermal dependence of the isothermal magnetic entropy change $\Delta S_M$ was calculated from magnetization isotherms recorded with increasing fields by numerical integration of one of the thermodynamic Maxwell's relations $\Delta S_M(T) = \mu_0 \int_0^H \left( \frac{\partial M(H,T)}{\partial T} \right)_H dH$ as described in [40]. The $-\Delta S_M(T)$ of each compound studied was evaluated in the vicinity of the magnetic transitions for field changes $\mu_0 \Delta H$ up to 5 T, with field steps of 0.2 T, and temperature increments of 2 K.

**3-Results and discussion**

**3.1 Structures at 300 K**

$Y_{0.9}Gd_{0.1}Fe_2H_{1.75}$ crystallizes in a cubic structure with superstructures refined by a doubling of the cubic cell parameter as for the non Gd-substituted $YFe_2H_{1.75}$ hydride (*I-43m* space group, and $a$ = 15.3659(1) Å). $Y_{0.9}Gd_{0.1}Fe_2H_{2.9}$ contains 92 % of a cubic MgCu$_2$ phase (*Fd-3m* S.G., $a$ = 7.8551(1) Å) and 8 % of monoclinic phase (*C2/m*, *M1*). The structure and cell parameters of the six hydrides (*C1, M1, C2, M2, C3, O*) which were previously studied in [36], are also reported in supplementary material (Table SI1).

In Fig. 1, the evolution of the relative cell volume variations ($\Delta V/V$) of the different $Y_{0.9}Gd_{0.1}Fe_2H_x$ hydrides versus H content is compared to those of $YFe_2$ deuterides, which were previously studied by neutron diffraction. The cell volume expansion reaches 29 % for the maximum H content. The $Y_{0.9}Gd_{0.1}Fe_2H_x$ hydrides and $YFe_2D_x$ deuterides present a very similar evolution with a deviation from linearity and a tendency to saturate above 3.5 H(D)/f.u.. The larger cell volume increase of $Y_{0.9}Gd_{0.1}Fe_2H_x$ hydrides compared to $YFe_2D_x$ deuterides at high H(D) content can be attributed to the H(D) isotope effect.

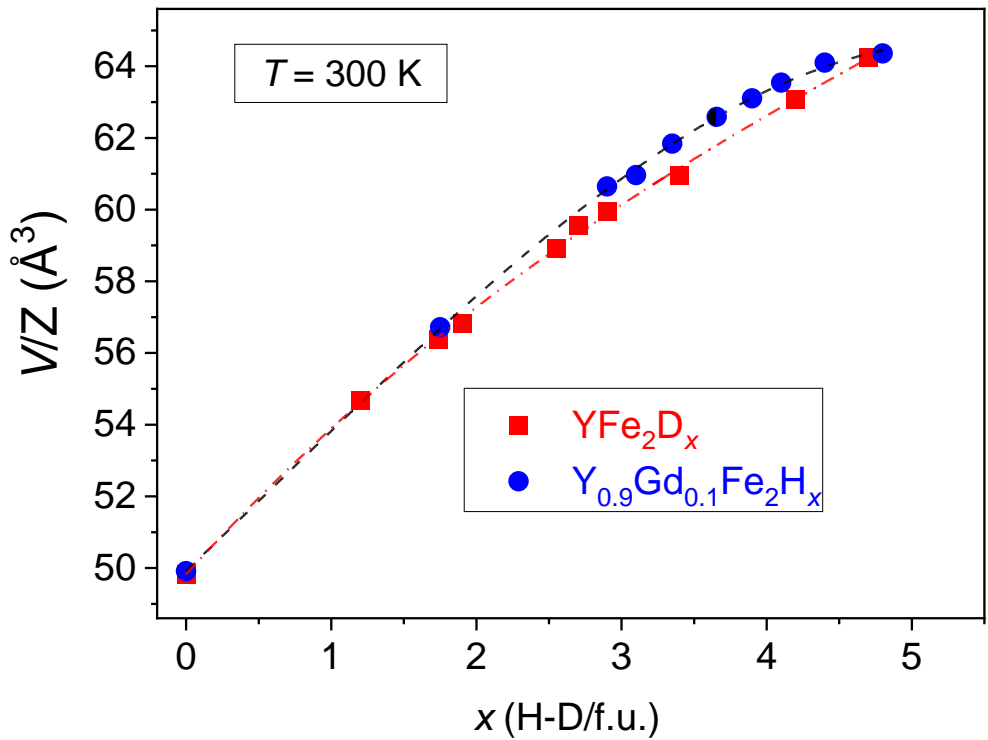

**Fig. 1**. Cell volume swelling of $Y_{0.9}Gd_{0.1}Fe_2H_x$ and $YFe_2D_x$ compounds at 300 K. The error bars are smaller than the symbols. The curves have been fitted by a polynomial function.

### 3.2 Magnetic properties

The $M_{5K}(H)$ magnetization curves at 5 K of selected $Y_{0.9}Gd_{0.1}Fe_2H_x$ hydrides (Fig.2(a)) are characteristic of a ferromagnetic or ferrimagnetic behavior. The–magnetization versus H-concentration at 5 K decreases slightly between 3 and 3.9 H/f.u. (-0.3(1) $\mu_B$/H atom), then with a larger slope above 4 H/f.u. (-3.8(2) µ$_B$/H atom) (Fig. 2(b)). This sharp magnetization decreases above 4 H/f.u. was also observed for YFe$_2$D$_x$ compounds. According to DFT calculations, the cell volume expansion tends to localize the Fe 3d band and to increase the Fe moment, but for a critical H concentration near 4 H/f.u. the strong Fe-H bonding lowers the Fe-Fe interactions [26, 27, 41]. As Gd is a heavy rare earth, an antiparallel orientation of the Fe and Gd magnetic sublattices reduces the saturation magnetization. A reduction of 0.60(2) µ$_B$ was observed for Y$_{0.9}$Gd$_{0.1}$Fe$_2$D$_{4.2}$ compared to corresponding YFe$_2$D$_{4.2}$ [33]. Magnetization isotherms of four selected samples at temperature between 2 and 320 K are presented in Fig. SI1 (supplementary materials).

The temperature dependence of $M_{0.03T}(T)$ magnetization of selected hydrides measured in the field of 0.03 T are reported in Fig. 2(c). The monoclinic *M1* and cubic *C2* hydrides both display a sharp transition around $T_C$ = 300 K with a thermal hysteresis of about 5 K. A shape difference between the two $M_{0.03T}(T)$ curves is however visible between 150 K and 300 K, as the magnetization curve of the *M1* compound becomes lower than the *C2* one and the transition towards a paramagnetic state more progressive. A small difference is also visible at low field in the $M_{5K}(H)$ curves, with a faster saturation for *C2* compared to *M1*.

The compounds containing the *M2* phase show a sharp magnetization decrease at 144 K, which can be attributed to a reversible FM-AFM transition as in the YFe$_2$(H, D)$_{4.2}$ compounds The AFM cell is doubled along the *b*-axis compared to the FM cell. Its magnetic structure is described by a stacking of Fe ferromagnetic layers oriented in antiparallel directions and separated by an intermediate non-magnetic Fe layer in the (*a*, *c*) plane. In both FM and AFM magnetic sublattices the Fe moments remain parallel to (*a*, *c*) plane, i;e. the perpendicular to the *b*-axis [30]. The FM-AFM transition itself is driven by the loss of magnetic moment of the Fe atoms belonging to the intermediate Fe layer. These Fe atoms are surrounded by a larger quantity of H(D) neighbors than the others Fe atoms, explaining their IEM behavior by the more numerous Fe-H bonds [30]. The Néel temperature $T_N$ for this phase is 200 K and corresponds to the temperature at which each ferromagnetic layer becomes paramagnetic. The compensation between the antiferromagnetically coupled ferromagnetic Fe layers, explain why no magnetic change is observed at $T_N$, whereas it is observed in cell volume variation.

This transition at 144 K is not observed for the sample containing a mixture of *C3* and *O* phases, whereas a sharp increase of the magnetization is observed below 15 K and 10 K for the samples containing the *C3* phase (inset of Fig. 2(c)). It is probably related to Gd ordering at low temperature as it was observed in ErFe$_2$D$_5$. Above 150 K, a larger value of the magnetization is observed for samples containing both *C3* and *O* phases compared to the *M2* one, and it increase with the weight fraction of *C3* phase. As the FM-AFM transition of the *M2* phase is related to its monoclinic structure, the larger magnetization for the compound containing the cubic *C3* phase indicates that without the monoclinic distortion, a weak ferromagnetic behavior exists up to room temperature.

The evolution of the transition temperatures versus H concentration shows a discontinuous variation (Fig. 2d), with $T_C$ near 300 K for $x < 4$, $T_{FM-AFM}$ at 144 K for $4 < x < 4.3$ and a magnetization increase below 10 K.

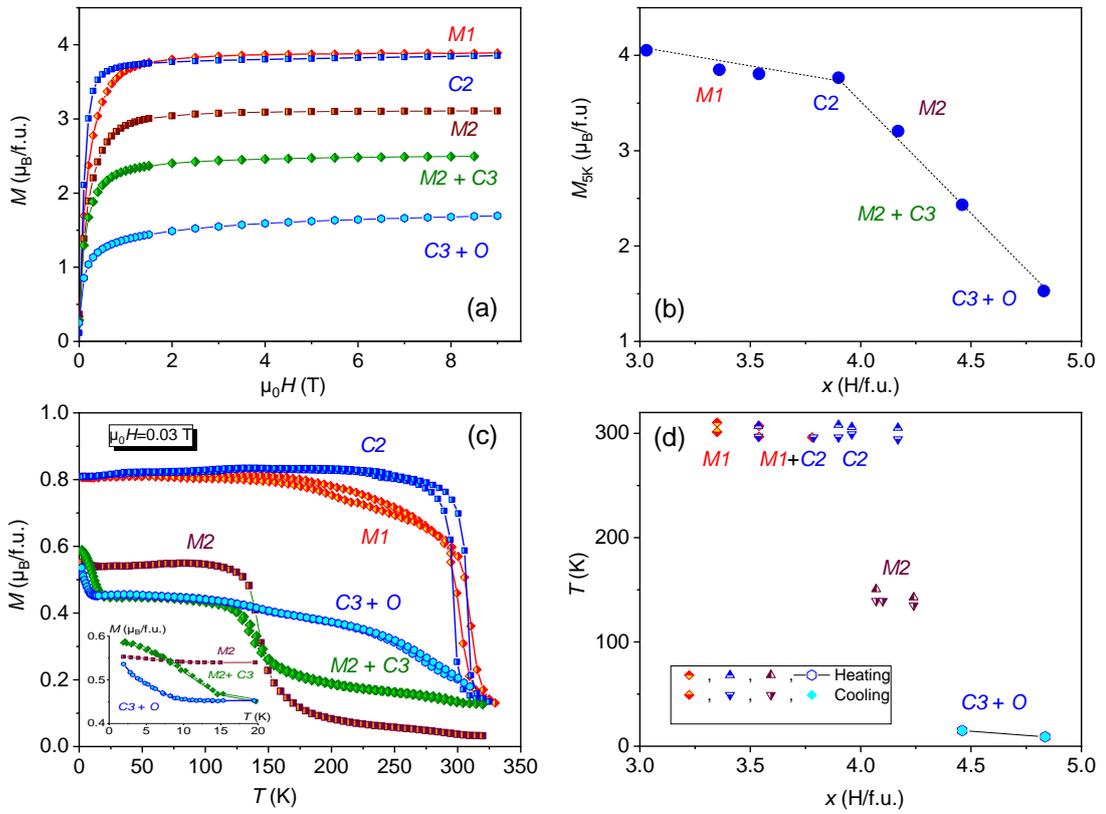

**Fig. 2**: (a) $M_{5K}(B)$ and (b) magnetization at 5 K, (c) $M_{0.03T}(T)$ and (d) magnetic transition temperatures versus H concentration of $Y_{0.9}Gd_{0.1}Fe_2H_x$ hydrides. Two samples presented in (a) and (c) contain a mixture of two phases: 31 % *M2*+ 69% *C3* (*M2+C3*) and 20% *C3*+80% *O* (*C3+O*). Inset (c): zoom below 20 K.

### 3.3 Structural properties

The *M1* and *M2* hydrides both crystallize in a monoclinic structure, which cell parameters evolution is compared in Fig. 3. All the SR-XRD patterns were refined below their order-disorder transition at $T_{O-D}$ in a monoclinic *C2/m* space group. Considering only the lowering of crystal symmetry from $Fd\text{-}3m$ (parent compound) to $C2/m$ space group, the amplitude of the distortion can be estimated by reporting the $a_M$, $b_M$ and $c_M$ cell parameters ($p$) described in an equivalent cubic lattice ($a_M.\sqrt{(2/3)}$, $b_M.\sqrt{(2)}$ and $c_M.\sqrt{(2)}$) (Fig. 3(a) and (d)). Compared to a non-distorted cubic cell, both hydrides display common features: a contraction of the $a_M$ and $c_M$ parameters (basal plane) and an expansion of the $b_M$ monoclinic parameter [36]. In addition, the monoclinic angle $\beta_M$ is smaller than 125.265°, the calculated value for the non-distorted cell (Fig. 3(b) and (e)). This structural distortion is attributed to the long-range order of the interstitial hydrogen atoms in tetrahedral $AB_2$ and $AB_3$ sites. As explained in [42] the long-range hydrogen order in Laves phase hydrides results from both the repulsive interaction between H atoms and the partial H occupation into all the possible sites. At $T_{O\text{-}D}$, the short-range order is maintained, but the long-range order disappear above a distance corresponding to the size of the cell.

Nonetheless, the details of the thermal evolution of $M1$ and $M2$ phases are clearly different and can be correlated to the magnetic properties previously described in section 3.2.

The thermal variations of the cell parameters belonging to the $M1$ phase are presented in Fig. 3 (a) for $a_{M1}$, $b_{M1}$ and $c_{M1}$, (b) for the monoclinic $\beta_{M1}$ angle and (c) for the reduced cell volume $V_{M1}/Z$ with Z = number of f.u./cell. The $a_{M1}$, $c_{M1}$ and $\beta_{M1}$ monoclinic cell parameters decrease upon heating down to a minimum at $T_C$ = 310 K and then increase again up to 350 K. The $b_{M1}$ cell parameter shows a different behaviour: it slightly increases up to 275 K and then decreases from 275 K to 350 K. Above 350 K, an O-D disorder transition towards a disordered cubic phase is observed as described in [36]. The small expansion of $b_{M1}$ upon heating up to 275 K is related to the thermal expansion and the reduction of the distortion. The opposite variation of the cell parameters between $T_C$ and $T_{O\text{-}D}$ is rather due to a progressive reduction of the monoclinic distortion which precedes the O-D transition and can be attributed to displacements of hydrogen atoms.

The cell volume $V_M$ shows a small bump around 125 K and a minimum around 310-320 K. The cell volume contraction between 120 and 310 K is $\Delta V/V_0$ = 1.9 $10^{-3}$ with $V_0$ taken at $T_C$. This value is slightly smaller than for isostructural hydride $Y_{0.9}Pr_{0.1}Fe_2D_{3.5}$ in the same temperature range ($\Delta V/V_0$ = 3.0 $10^{-3}$) [43]. A simple extrapolation at low temperature of the cell volumes in both paramagnetic and ferromagnetic ranges yields $\Delta V/V \approx 4.4\ 10^{-3}$, which is the same order of magnitude than for $YFe_2$ ($\omega_s$ = 2.2 $10^{-3}$ at 4.2 K) and $LuFe_2$ ($\omega_s$ = 4.1 $10^{-3}$ at 4.2 K) [44]. Between 120 K and 310 K, the relative cell parameter variations ($\Delta p/p$) normalized to the value at $T_C$ (inserted in Fig. 3a) shows that the relative contraction of $a_{M1}$ and $c_{M1}$ is almost 4 time larger than the expansion of $b_{M1}$. An extrapolation of the cell parameters variation to low temperature is difficult as above $T_C$ the cell parameter variation is due to both thermal expansion and the reduction of the monoclinic distortion. Therefore, only the variation between 120 K and $T_C$ will be considered and the order of magnitude discussed. The anisotropic cell parameter variation of $Y_{0.9}Gd_{0.1}Fe_2D_{3.35}$ below $T_C$ $\Delta a/a_0 = 2.7\ 10^{-3}$ and $\Delta c/c_0 = 1.6\ 10^{-3}$ are significantly larger than in the reference alloys $YFe_2$ ($\lambda_{111} < 10^{-5}$) and $GdFe_2$ ($\lambda_{001} < 10^{-5}$) and comparable to that observed for $DyFe_2$ ($\lambda_{111} < 3.10^{-3}$). The negative value obtained for $\Delta b/b_0 = -0.5\ 10^{-3}$ between 120 K and $T_C$ is even closer to the value found for $HoFe_2$ ($\lambda_{100} = -0.75\ 10^{-3}$) than for $DyFe_2$ ($\lambda_{100} = -0.07\ 10^{-3}$). This indicates that the $a_{M1}$ and $c_{M1}$ variations, opposite to the expected thermal expansion, can be attributed to a spontaneous anisotropic magnetostriction effect as already observed in isostructural $Y_{0.9}Pr_{0.1}Fe_2D_{3.5}$ deuteride [43] and in other $R$Fe$_2$ compounds [44]. The magnetostriction in $R$Fe$_2$ compounds has been attributed to the electrostatic interaction between the anisotropic 4f electron shell of the $R$ ion and the crystal field [45]. As the magnetocristalline anisotropy of Y and Gd moments is very small, the contribution of the Fe magnetic sublattice to the spontaneous magnetostriction in $Y_{0.9}Gd_{0.1}Fe_2H_{3.35}$ hydride should therefore be considered. This is probably related to the enhancement of the itinerant character of Fe moment due to the strong Fe-H bonding. Indeed, if the Fe contribution to the magnetocrystalline anisotropy remains weak in $R$Fe$_2$ compounds which are strong ferromagnets, the influence of hydrogen insertion on their magnetic properties with a decrease of Fe-Fe interactions and a behaviour closer to that of $R_2Fe_{17}$ compounds when the H content increases cannot be neglected [29]. For example, in $Y_2Fe_{17}$, a spontaneous magnetostriction of $\lambda_a = 1.5\ 10^{-3}$ and $\lambda_c = 8.04\ 10^{-3}$ due to Fe has been observed at 5 K. In [43], the contraction of the $a_{M1}$ and $c_{M1}$ cell parameters upon heating was related to the reduction of the Fe moments oriented parallel to the $(a, c)$ plane, in agreement with critical exponent analysis showing a good agreement with a 3D XY model. A similar cell parameter variation was observed for $Y_{0.9}Pr_{0.1}Fe_2D_{3.5}$ with minima of $a_M$ and $c_M$ parameters at $T_C = 274$ K [43]. It is also interesting to notice that the Curie temperature decreases upon $R$ for Y substitution from 345 K for $YFe_2D_{3.5}$ [46] to 310 K for $Y_{0.9}Gd_{0.1}Fe_2H_{3.35}$ and 274 K for $Y_{0.9}Pr_{0.1}Fe_2D_{3.5}$.

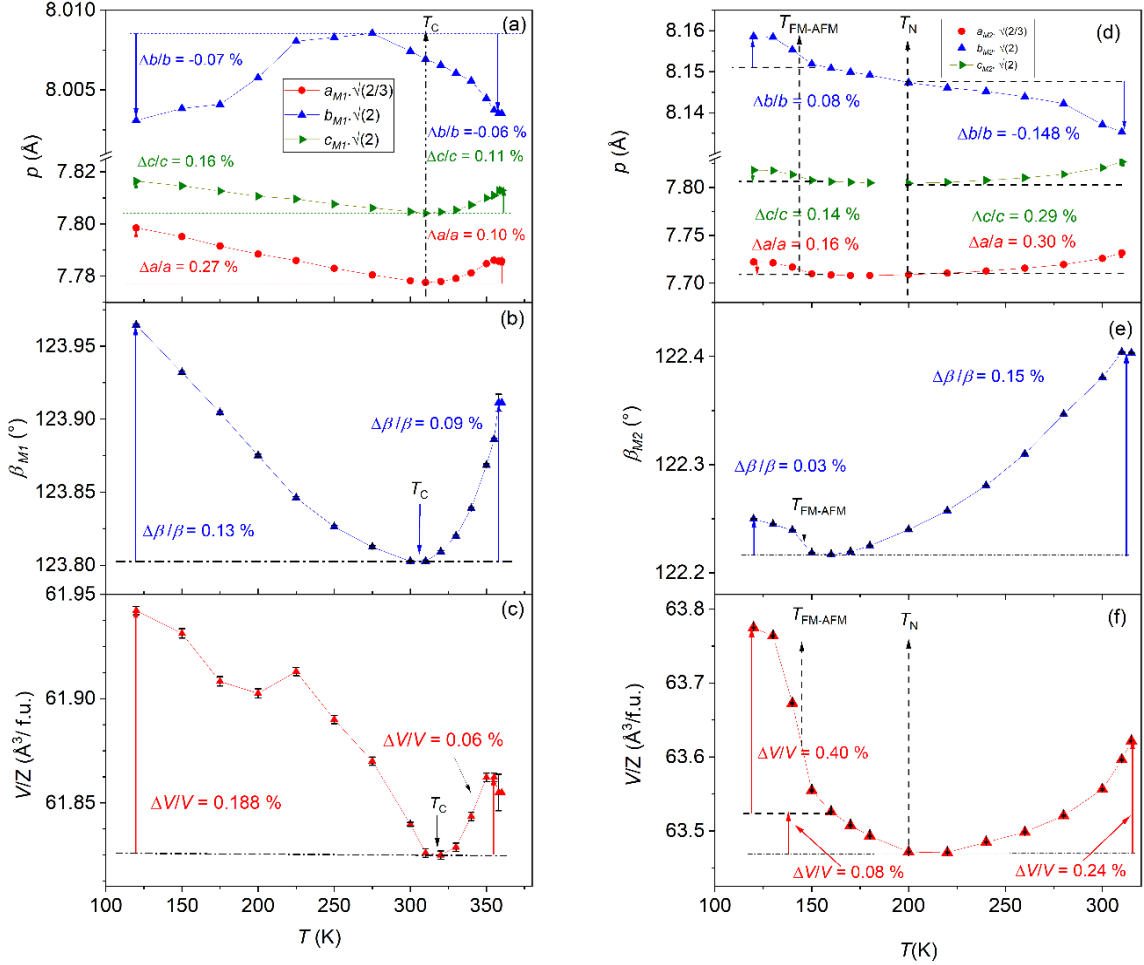

**Fig. 3**: Cell parameters of the monoclinic *M1* (left) and *M2* (right) hydrides. Details are given in the text.

The cell parameters of the monoclinic *M2* phase are displayed in the right part of Fig. 3. The $a_{M2}$, $b_{M2}$ and $c_{M2}$ cell parameters are described in an equivalent cubic lattice in Fig. 3 (d), the monoclinic $\beta_{M2}$ angle in Fig. 3(e) and the reduced cell volume $V_{M2}/Z$ in Fig. 3 (f).

All cell parameters show a sharp decrease between 130 and 160 K which corresponds to the FM-AFM transition observed by magnetic measurements. The cell volume jump ΔV/V of 0.4 %, close to that observed for YFe$_2$H$_{4.2}$ (0.43 %) [30], confirms the first order character of this transition and is characteristic of the itinerant electron metamagnetic behaviour of the Fe sublattice. The AFM structure, which is doubled along the ***b*** axis compared to the FM structure, can be described by a stacking of ferromagnetic Fe layers along the ***b***-axis which are coupled antiferromagnetically to each other. The FM-AFM transition was explained by the loss of the Fe moment belonging to one Fe site lying in the middle plane of the AFM cell leading to an antiparallel orientation of the Fe moments above and below this plane.

The $a_{M2}$ and $c_{M2}$ relative reduction are of 0.16 and 0.14 % respectively, whereas the $b_{M2}$ relative reduction is 0.08 %, i.e. two time smaller. As previously explained, this difference of cell parameter variation was already observed for isostructural $YFe_2(H_\alpha D_{1-\alpha})_{4.2}$ compounds [30]. The NPD analysis showed that the Fe moments were oriented parallel to the (*a, c*) plane in both FM and AFM structures and that the cell contraction was larger in the basal plane than along the ***b*** axis [30] as also observed for $Y_{0.9}Gd_{0.1}Fe_2H_{3.35}$.

From 150 to 200 K the cell volume continues to slightly decrease, then increases again above. Considering the NPD results on $YFe_2H_{4.2}$ and $YFe_2D_{4.2}$, the change of slope at 200 K can be attributed to the AFM-PM transition at $T_N$. Despite the volume effect is less pronounced, the reduction of the Fe moment in each ferromagnetic layer also reduces the repulsive exchange interactions between the Fe moments, and therefore the overall volume. Then above $T_N$, $a_{M2}$, $c_{M2}$, $\beta_{M2}$ parameters and $V_{M2}/Z$ increase again upon heating whereas the $b_{M2}$ parameter decreases, up to 310 K. As for the *M1* hydride, this evolution results from the thermal expansion and the reduction of the monoclinic distortion which is additive for $a_{M2}$, $c_{M2}$ and $\beta_{M2}$ and subtractive for $b_{M2}$. As the expansion is around 0.295 % for the *a* and *c* parameters and -0.148 % for the $b_{M2}$ parameter between 200 and 310 K, it is possible to estimate that the pure thermal expansion is of 0.074 % in this temperature range.

The SR-XRD patterns of a sample containing a mixture of *C2* and *M1* phases has been measured upon heating between 200 and 360 K. The weight fraction and cell parameter variation of the *C2* and *M1* phases are presented in Fig. 4. For clarity, in the following *M1*-I corresponds to the monoclinic phase presented in Fig.3 (left) and *M1*-II for the *M1* phase displayed in Fig. 4. Note that these two *M1* phases can be described in the same monoclinic space group (*C2/m*) and differ mainly by their H concentration.

The weight fraction of the *M1*-II phase increases from 16 to 40 % between 200 and 315 K at the expense of the cubic *C2* phase and decreases above 315 K. The cell volume of M1-II increases by 0.54 % between 200 and 315 K, whereas that of *C2* increases by 0.12 % from 200 K to 270 K and decreases by 0.3 % from 270 K to 360 K. The thermal behaviour of the *M1*-II phase is clearly different from the *M1*-I one, where a minimum instead of a maximum of *V*/Z is observed near $T_C$. The change of weight fraction of each phases indicates that the main mechanism below 315 K should be a transfer of hydrogen from *C2* to *M1*. The cell volume variation versus temperature for *M1*-II and *C2* phases results from a competition between thermal expansion, magnetovolumic effect and H diffusion between these two phases. Considering, the cell volume variation of *M1*-I between 200 and 315 K due to thermal expansion and magnetovolumic effect, the cell volume increases due to H transfer is of 0.404(1) Å. In this H concentration range the cell volume expansion caused by H atom is around 2.3 H/Å$^3$, which means that about 0.17(2) H/f.u. is transferred from *C2* to *M1*-II between 200 K and 315 K. Above 320 K the weight fraction of *M1*-II lowers compared to the cubic phase and both phases are paramagnetic. This change can be attributed either to a H transfer to the opposite direction (*M1*-II towards *C2*) or to the onset of an O-D transition from the ordered towards the disordered cubic phase. The analysis of the cell parameter variation (Fig. SI2, supplementary materials) of both monoclinic *M1*-I and *M1*-II phases shows that: i) at 200 K, the *b* cell parameter of phase *M1*-II is larger than the one of phase *M1*-II, whereas *b* and *c* are similar in both phases, ii) the thee *a*, *b* and *c* parameters of phase *M1*-II expands up to 320 K and decreases above 320 K whereas a reduction of $b_{M1\text{-}I}$ and an expansion of $a_{M1\text{-}I}$ and $c_{M1\text{-}I}$ is observed above 300 K, iii) the monoclinic angle *β* is larger for the *M1*-II phases below 350 K, but a minimum of *β* is observed for both compounds at $T_C$. These data can be interpreted as follow: i) the monoclinic distortion is larger for the *M1*-II phase with an expansion along the ***b*** axis, ii) for *M1*-I, a reduction of the monoclinic distortion is observed above $T_C$, whereas for *M1*-II there is a competition between H transfer from *M1* towards *C2* and a reduction of the monoclinic distortion when approaching the O-D transition.

The *C2* phase SR-XRD pattern display superstructure peaks indexed in a double cubic cell and which intensities decrease upon heating. A zoom of XRD patterns between 200 and 360 K is presented in Fig. SI3 in supplementary materials. Some superstructure peaks disappear above 325 K (Fig. SI4, supplementary materials), in the paramagnetic range. This can also correspond to the O-D transition for this *C2* phase with the loss of long-range order.

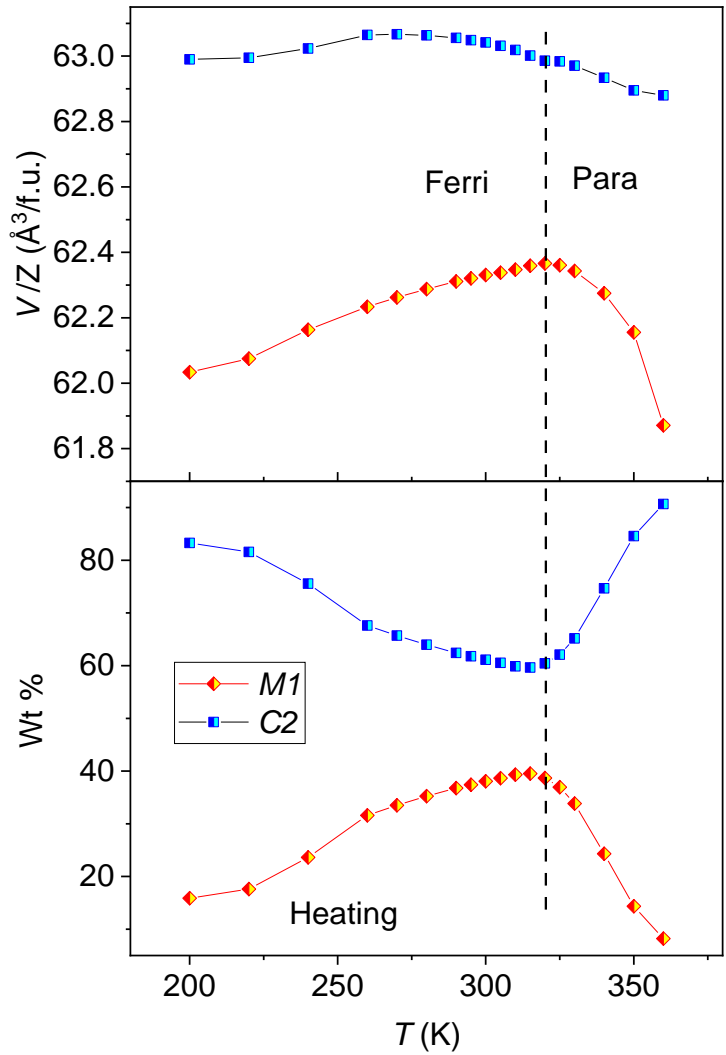

**Fig. 4**: Cell volume and weight fraction upon heating of the monoclinic *M1* and *C2* hydrides in a compound containing a mixture of *M1* and *C2* phases.

As the hydrogen concentration increases, the number of H neighbours around each Fe atom becomes larger and modifies the Fe-Fe interactions. For $x = 4.2$, each Fe is surrounded by 4 to 5 H neighbours and the Fe sublattice presents an itinerant electron metamagnetic behaviour whereas for $x > 4.8$ the Fe sublattice become paramagnetic.

### 3.4 Structural and magnetic phase diagram

The evolution of the cell volume below $T_{O-D}$ of all the $Y_{0.9}Gd_{0.1}Fe_2H_x$ compounds that have been studied by SR-XRD are plotted in Fig. 5. The cell volume changes due to hydrogen order-disorder have not been reported in this graph as they were already detailed in [36].

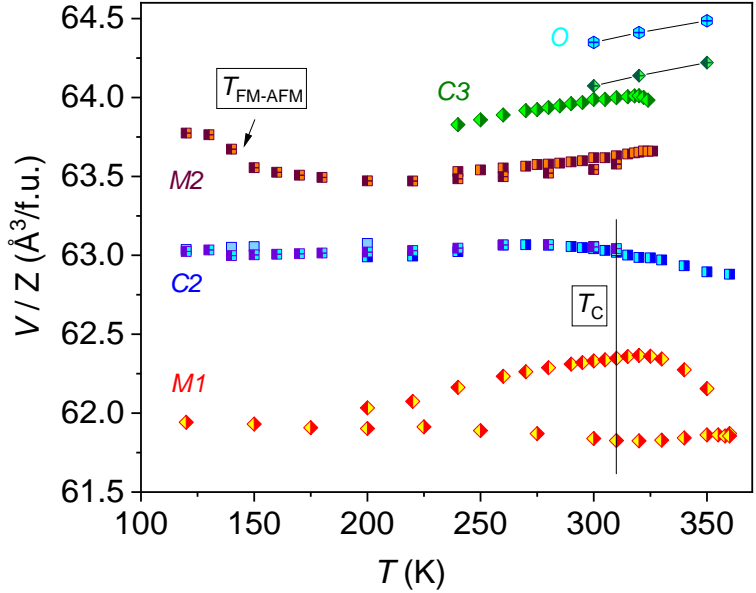

**Fig. 5**: Cell volume evolution versus temperature of the different phases below the O-D transitions measured for several samples. The magnetic transitions temperatures ($T_{FM-AFM}$ and $T_C$) are indicated in box. The two different $V/Z$ variations of *M1*, reflect the large solubility range: the lower one corresponds to a single *M1* phase whereas the upper one was measured for a mixture of *M1* + *C2*.

For *M1*, *C2*, *M2* and *C3* phases the results obtained on different samples are combined. For *M1*, two different variations were observed, the one with smaller volume is related to $Y_{0.9}Gd_{0.1}Fe_2H_{3.35}$ hydride which is described in detail in Fig. 3 left. The second one was measured for a sample which contains a mixture of *M1* + *C2*. This shows that M1 exist in a broad range of H content. The evolution of *C2*, was obtained from 3 different samples (*M1* + *C2*, *C2*, *C2* + *M2*) and shows that it exists only for a narrow H concentration range. The evolution of *M2* cell volume corresponds to the sample presented in Fig. 3 and a second sample containing *M2* and *C3* and shows also that it forms only for a narrow H content. Two set of data were obtained for *C3* and measured in a smaller temperature range, one related to a mixture of *M2*+*C3* and the second one to *C3*+*O*. Here also, we observe that this phase form for a relatively narrow H content. For the orthorhombic phase, only few patterns were obtained, and desorption occurs above 350 K.

The structural and magnetic results described in this work are combined with those describing the O-D transitions on the same samples in [36] to draw a structural and magnetic phase diagram (Fig. 6). This indicates that except the *M1* phase the other hydrides exists in a narrow temperature range and that all the magnetic transitions are found systematically below the O-D structural transitions.

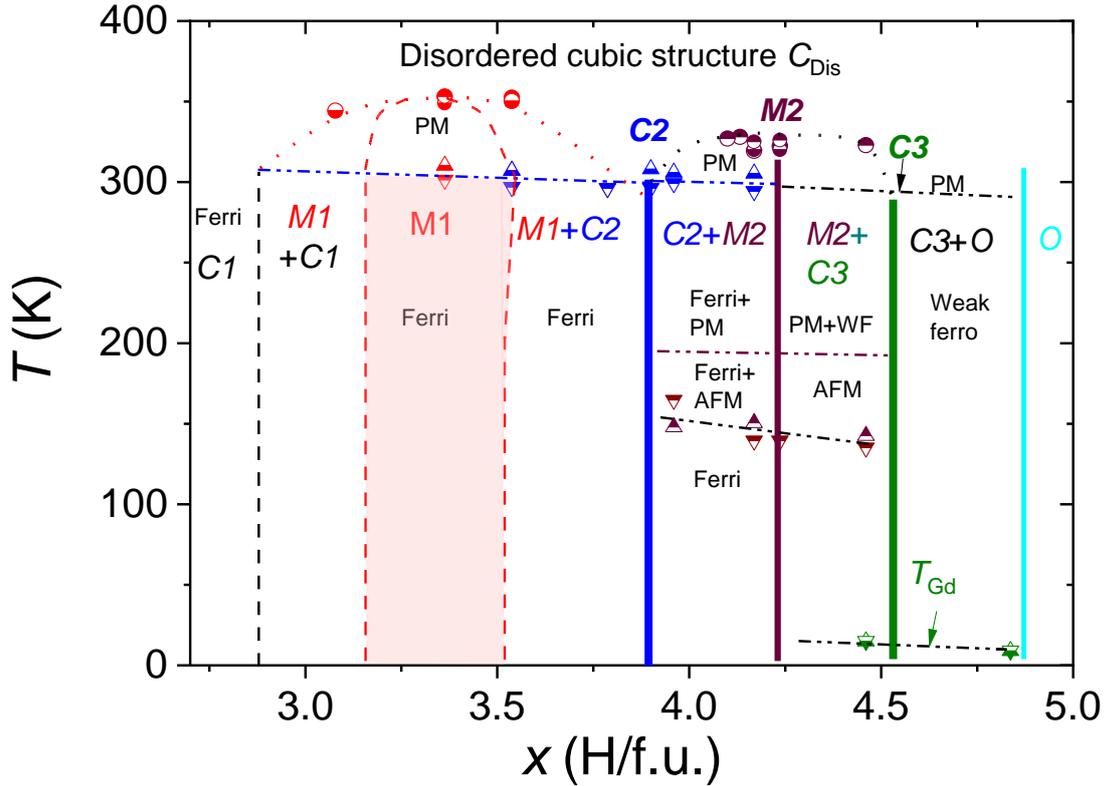

**Fig. 6**: Schematic structural and magnetic phase diagram of $Y_{0.9}Gd_{0.1}Fe_2H_x$ hydrides with *C1* (black), *M1* (red), *C2* (blue), *M2* (brown), *C3* (olive) and *O* (cyan) phases. The magnetic transitions are indicated with triangles (up: heating and down: cooling) whereas the O-D transitions are indicated by circles (◖: heating and ◗ cooling).

### 3.5 Magnetocaloric properties

The magnetic entropy variations measured for four samples were obtained by applying the Maxwell equation to the $M_T(B)$ curves. The data in Fig. 7 were obtained with a field variation between 0 and 5 T.

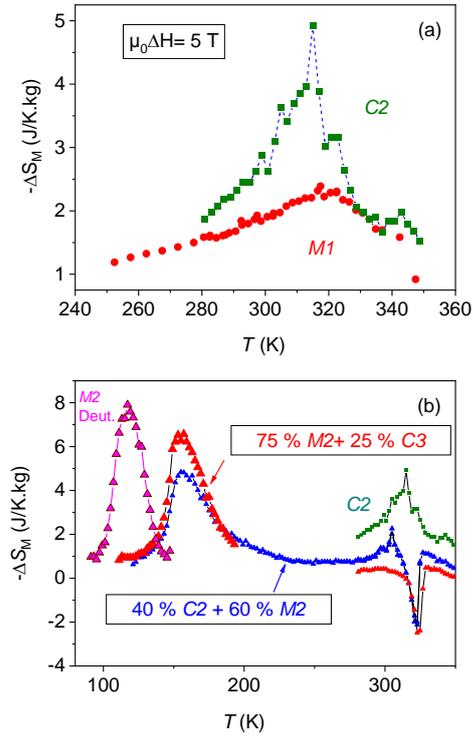

**Fig. 7**: Magnetic entropy variation $\Delta S_M$ versus temperature for different hydrides with a field variation $\mu_0 \Delta H$ =5 T. (a) for *M1* and *C2* single phases and (b) for samples containing *M2* including a monoclinic $Y_{0.9}Gd_{0.1}Fe_2D_{4.2}$ deuteride which is added for comparison.

On the left part, the magnetic entropy variation ($\Delta S_M$) curves of the *M1* and *C2* phases (Fig. 7(a)) are compared. Both present a negative $\Delta S_M$ peak at 320 and 310 K respectively corresponding to $T_C$. The maximum of $\Delta S_M$ is more than two times larger for the *C2* phase ($\Delta S_M$ = -5 J/ K.kg) compared to the *M1* phase ($\Delta S_M$ = -2 J/ K.kg). This can be related to the different $M(T)$ behavior near $T_C$, as the magnetization decrease is sharper for *C2* compared to *M1*. This is due to the structural anisotropy of the monoclinic phase compared to the cubic one. The $\Delta S_M$ values are small, because of the second order of the transition and the absence of the cell volume variation at the transition.

The $\Delta S_M$ curves of compounds presenting a mixture of *M2* + *C2* and *M2* + *C3* phases are presented in Fig. 7(b). The large negative peak at 155 K, can be attributed to the first order FM-AFM transition in the *M2* phase. Its maximum increases with the *M2* percentage ($\Delta S_M$ = - 6.5 J/K.kg and -4.9 J/K.kg respectively). Considering the weight fraction of *M2* and the absence of magnetic transitions for *C2* and *C3* phases at 144 K, a magnetic entropy variation $\Delta S_M$ = -8(2) J/K.kg is obtained for *M2*, which compares well with the value obtained for the corresponding $Y_{0.9}Gd_{0.1}Fe_2D_{4.2}$ deuteride ($\Delta S_M$ = -8(2) J/K.kg). This magnetic entropy variation is larger than for *M1* and *C2*, as it corresponds to a first order transition associated to a cell volume variation. A positive sharp peak is also observed at 323 K for both hydrides. It corresponds to an inverse MCE effect and has about the same amplitude than the *C2* peak but with an opposite sign. A close examination of the figure 7(b), shows that the compound containing a mixture of *C2* and *M2*, also displays the beginning of a $\Delta S_M$ negative peak a 305 K, before the positive one at 323 K. As it is not present in the sample containing *M2* and *C3* phases, this $\Delta S_M$ peak belongs therefore to the *C2* phase, and it confirms that the contribution related to magnetic and hydrogen O-D are of opposite signs. Similar effects have been observed at $T_{O-D}$ for $Y_{0.9}Pr_{0.1}Fe_2D_{3.5}$ [38] and the positive $\Delta S_M$ associated to hydrogen O-D transition was explained by the contribution of the lattice entropy variation ($\Delta S_L$) to the total entropy variation $\Delta S$.

## 4. Conclusions

The influence of the hydrogen insertion on the structural, magnetic and magnetocaloric properties of the $Y_{0.9}Gd_{0.1}Fe_2$ compound has been investigated to solve the structural and magnetic phase diagram for hydrogen rich compounds and obtain magnetic transitions and related MCE near room temperature. Hydrogen insertion induces a lowering of crystal symmetry and 6 different phases (*C1*, *M1*, *C2*, *M2*, *C3* and *O*) with different crystal structures (*C* for cubic, *M* for Monoclinic and *O* for orthorhombic) were observed below $T_{O-D}$. Anisotropic cell parameter variations were observed for the two monoclinic *M1* and *M2* phases. The *M1* phase exists in a large H concentration range, whereas the *C2* phase forms in a narrow concentration range. It has been observed that the H concentration between *M1* and *C2* phases depends on temperature. The three other phases (*M2*, *C3* and *O*) are also observed for a narrow H range. Upon H insertion, a discontinuity of the total magnetization is observed above 4 H/f.u., it can be related to the strong Fe-H bonding which decreases the Fe-Fe interaction, opposite to the cell volume expansion. A step decrease of the magnetic ordering temperature is also observed versus H concentration, which can be related to both structural and electronic effects. The *M1* and *M2* phases both display spontaneous anisotropic magnetostriction effects around $T_C$ = 310 K for *M1* and $T_{FM-AFM}$ = 144 K for *M2* related to the preferential orientation of the Fe moments in the (*a*, *c*) plane. A combined structural and magnetic phase diagram has been proposed. A significant magnetic entropy variation is found for *M2* near $T_{FM-AFM}$. Weaker $\Delta S_M$ peaks are observed near 310 K for the *M1* and *C2* phases. The *M1*, *C2* and *M2* phases show order-disorder transitions related to displacements of metallic and hydrogen atoms. An inverse MCE is observed near $T_{O-D}$ for the sample containing the *M2* phase. It can be attributed to the contribution of the lattice entropy variation of the total entropy.

To conclude, it is possible to adjust the magnetic ordering temperature near room temperature by playing with the H content, but the MCE remains quite weak, due to second order type transitions.

**CRediT authorship contribution statement**

V. Paul-Boncour: Conceptualization, Formal analysis, Investigation, Methodology, Project administration, Supervision, Validation, Visualization, Writing – original draft, Writing – review & editing. K. Provost: Investigation. E.Alleno: Investigation, Writing – review & editing. , T. Mazet: Investigation, Writing – review & editing, A. N'Diaye: Investigation, Formal analysis, Writing – review & editing, F. Couturas: Investigation.

**Declaration of Competing Interest**

The authors declare that they have no known competing financial interests or personal relationships that could have appeared to influence the work reported in this paper.

**Acknowledgments**

We are thankful to the synchrotron SOLEIL for the allocated beam time on the CRISTAL beam line (proposal 20160868), and particularly to Erik Elkaim as local contact. We also acknowledge K. Hakkaoui for some sample preparation, C. Rakotoarimanana for her participation to the synchrotron experiments, H. Ghulam and H. Karami for SR-XRD data analysis during their trainship.